\documentclass[aps,prl,twocolumn,superscriptaddress,fleqn,a4paper]{revtex4-1}
\usepackage[ngerman,english]{babel}
\usepackage[utf8]{inputenc}   
\usepackage{xspace}
\usepackage{amsmath,tensor}
\usepackage{array,color}
\usepackage{relsize,exscale}  
\usepackage{textcase}	
\usepackage{soul,color}		
\usepackage{calc} 
\usepackage{makeidx,showidx}
\usepackage{graphicx}
\usepackage{wrapfig}
\usepackage{float}
\usepackage{textcomp}
\usepackage{siunitx}
\usepackage[version=3]{mhchem}
\usepackage{enumitem}
\usepackage{fancyref}
\usepackage[hidelinks=true]{hyperref}
\sethlcolor{green}
\setulcolor{red}
\usepackage{lineno}			

\newcommand{\erww} [1] {\ensuremath{\langle {#1} \rangle}}

\newcommand{\lsco} {{La$_{2-x}$Sr$_x$CuO$_4$}\@\xspace}

\newcommand{\hgryb} {{HgBa$_{2}$CuO$_{4+\delta}$}\@\xspace}

\newcommand{\ybco} {$\ce{YBa2Cu3O_{6+y}}$\@\xspace}

\newcommand{\ybcoE} {$\ce{YBa2Cu4O8}$\@\xspace}

\newcommand{\tc} {\ensuremath{T_{\rm c}}\@\xspace}
\newcommand{\cperp}{\ensuremath{c \bot B_0}\@\xspace}
\newcommand{\cpara}{\ensuremath{{c\parallel\xspace B_0}}\@\xspace}

\newcounter{exex}[section]

\makeatletter\newcommand\listofexamples{\section*{List of Examples}\@starttoc{xmp}}
\newcommand\l@example[2]{\par\noindent#1~\textit{#2}\par}
\makeatother






\makeatletter
\renewcommand\subsection{\@startsection 
	{subsection}{3}{0mm}
	{-\baselineskip}
	{0.5\baselineskip}
	{\centering \textbf }}
\makeatother

\makeatletter
\renewcommand\subsubsection{\@startsection 
	{subsubsection}{3}{0mm}
	{-\baselineskip}
	{0.5\baselineskip}
	{\centering  }}
\makeatother

\makeindex

\begin{document}




\title{A different NMR view of cuprate superconductors}

\date{\today}

\author{J\"urgen Haase}


\affiliation{%
Felix Bloch Institute for Solid State Physics, University of Leipzig,  Linn\'estr.\@ 5, 04103 Leipzig, Germany\\
}





\begin{abstract}
Nuclear magnetic resonance (NMR) is a powerful quantum probe, but the early conclusions on the physics of the cuprates, based on a limited set of data, have to be revised in view of recent findings and results from extensive literature analyses of most NMR data. These show two coupled electronic spin components that influence the nuclei, most easily seen with the planar Cu shift anisotropy. One component is spin from the recently identified ubiquitous metallic excitations, the other due to the intrinsic, antiferromagnetically coupled electronic Cu spin. Both components and its intricate interaction leave their imprint on nuclear shifts and relaxation. They also show a family dependence seen in the charge sharing between planar Cu and O. The main phenomena of the doping and temperature dependences of the interplay between both spins components are discussed in terms of an apparent phenomenology that awaits explanation from theory.
\end{abstract}


\maketitle

\subsection{The early NMR view}
With the advent of cuprate high-temperature superconductivity \cite{Bednorz1986} and the known power of nuclear magnetic resonance (NMR) in elucidating chemical and electronic properties of materials \cite{Slichter1990}, there was immediate interest, also from theory, in investigating the cuprates with NMR (for a review see \cite{Slichter2007}). The resonances from the Cu and O nuclei in the ubiquitous CuO$_2$ plane should contain essential information with the various nuclear transitions assigned to chemical sites, also from the charge reservoir layers. One has to remember that NMR of the planar sites concerns isotopes with larger spin, $^{63,65}$Cu ($I=3/2$) and $^{17}$O ($I=5/2$), so that the Zeeman resonance is split into $2I$ lines due to the electric hyperfine interaction. These splittings complicate the NMR spectra, but they also measure the local charge symmetry and we know, today, that the thus derived planar O hole content of the $2p_\sigma$ bonding orbital is perhaps the only normal state property that is directly related to the maximum critical temperature of superconductivity \cite{Rybicki2016,Kowalski2021}. 

It was quite natural that, in the early days, the focus was on comparing the spin shift and relaxation of doped cuprates to those of classical metals and superconductors, given that NMR can measure the electronic spin susceptibility \cite{Knight1949,Korringa1950,Schumacher1954} and delivered the first proof of BCS theory \cite{Hebel1957,Bardeen1957} with identifying the coherence peak in $^{27}$Al nuclear relaxation. With the cuprates being of type-II, even the investigation of the condensed state is possible as the magnetic field penetrates the sample and the residual diamagnetism in the large magnetic fields used for NMR is quite small. 

Indeed, first observations saw rapid nuclear relaxation and large shifts, both typical for the high density of states in metals, but also typical for more localized moments (and differences are difficult to detect with a local probe). Below the critical temperature of superconductivity ($T_\mathrm{c}$), shift and relaxation seemed to disappear, a clear indication of spin singlet pairing. Not anticipated was the finding first seen with the experiments on $^{89}$Y ($I=1/2$) of \ybco \cite{Alloul1989} (measurements only above \tc), that merely the strongly doped materials were metal-like in the sense that the NMR shift is temperature independent (Pauli spin susceptibility). As the doping decreases, the shifts begin to drop at increasingly higher temperatures compared to \tc. This was believed to be due to the opening of a spin gap (the pseudogap) at those temperatures for more local moments.
\begin{figure}
	\includegraphics[width=0.4\textwidth]{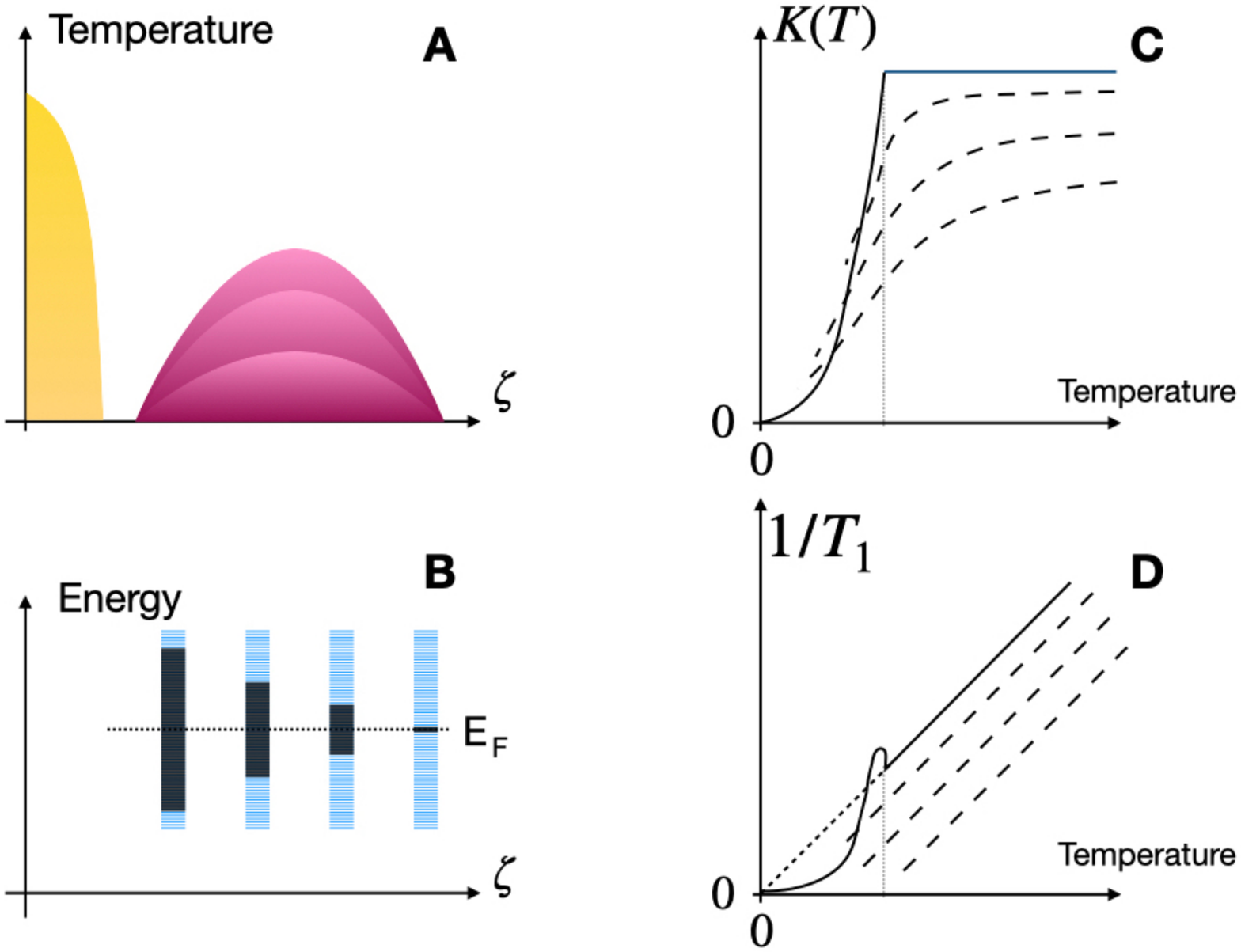}
	\caption{The traditional cuprate phase diagram (A); it does not distinguish between materials with different, highest $T_\mathrm{c}$. As a function of doping $\zeta$, (B) the peudogap closes in a material independent, metallic density of states. As a result (C), the Pauli susceptibility (full line), responsible for the Knight shift $K$ of a regular metal, undergoing a transition (dotted vertical line) to a spin singlet superconducting state, is suppessed and shows a temperature dependence due to the Fermi function (low energy states are missing); (D) this scenario leads to offsets in the Heitler-Teller relaxation rate (proportional to temperature) due to the missing low energy states (similar to the entropy).}
	\label{fig:fig1}
\end{figure}
Unfortunately, the early NMR experiments concerned mostly a few, available materials (\ybco, \ybcoE, \lsco). With the body of data still small, and despite a number of inconsistencies, generic features had to be assumed. For example, while the Cu magnetic shifts can be strongly temperature dependent (above and below \tc) when the magnetic field is perpendicular to the crystal $c$-axis (\cperp), they show no temperature dependence at all for \cpara for \lsco and \ybcoE, or only a rather weak one for \ybco. Since planar O shows strong temperature dependent shifts for all directions of measurement, as well, and they resemble those at planar Cu for \cperp, only an accidental cancellation of the hyperfine coupling for \cpara could explain the data \cite{Mila1989b} (in a single spin component model). 

Note that this means that the well-known, large negative hyperfine coefficient $A_\parallel$ from a hole in the $3d(x^2-y^2)$ orbital had to be cancelled by positive, transferred spin assumed to have an isotropic hyperfine coefficient $B$, i.e. $A_\parallel + 4 B \approx 0$. Note also that since $A_\perp$ is of much smaller magnitude than $A_\parallel$ \cite{Pennington1989}, only the transferred spin through $B$ survives in the description of the planar Cu shifts. Again, this removes a possible direct action of the onsite Cu spin and makes the Cu shift anisotropy irrelevant. Another important consequence of this assumption concerns relaxation. Fluctuating fields parallel to the crystal $c$-axis can only contribute to Cu relaxation if the former peak at the antiferromagnetic wave vector (to remove the constraint $A_\parallel + 4 B = 0$). Note, these were fluctuations that one could expect to be present. Finally, there is another important conclusion of this assumption. It defines the orbital shift as the shift observed at the lowest temperature for \cpara (which was by far too large, even larger than what one expects from an isolated Cu ion \cite{Pennington1989}).

For metals, the nuclear relaxation ($1/T_1$) is proportional to temperature ($T$), and $T_1T K^2=(\gamma_\mathrm{e}/\gamma_\mathrm{n})^2\hbar/(4\pi\mathrm{k_B})$, where $K$ is the Knight shift and $\gamma_n$ denotes the gyromagnetic ratios of electron and nucleus. This is the famous Korringa relation \cite{Korringa1950} that does not contain the notoriously unknown hyperfine coefficients. While this relation appeared to work to some extent for planar O and inter-planar Y, it failed most planar Cu data as the shifts were too small. So it was concluded that the Cu relaxation is enhanced from antiferromagnetic spin fluctuations that are shielded from planar O due to symmetry. Also, the shifts for planar Cu for \cperp are quite similar to those of planar O (that is involved in the spin transfer $B$), and this was seen as evidence for single component behavior of the cuprates, based on the data for two system, YBa$_2$Cu$_3$O$_{6.63}$ and \ybcoE \cite{Takigawa1991,Bankay1994}. 

While this accidental cancellation raised suspicions, there was no other sound explanation for the observations, and first principle calculations endorsed the hyperfine scenario, later \cite{Huesser2000}.\par\medskip

\subsection{More conflicting evidence}
We have been involved in extended tests of the NMR shift scenario, and more evidence turned up pointing to inconsistencies, for \lsco \cite{Haase2009b}, \ybcoE \cite{Meissner2011}, and \hgryb \cite{Haase2012,Rybicki2015}. For the latter family, there are large temperature dependent Cu shifts also for \cpara, but these are not even proportional to those measured for the other direction. This and the fact that the body of Cu and O NMR experiments on cuprates had grown tremendously over the years, due to diligent work by a number of groups, it seemed indicated to take a fresh look at the entire data, and some of us began with the collection of the planar Cu shifts \cite{Haase2017}, as one might think that the uniform response is rather simple. This turned out not to be true, and a complex shift phenomenology was formulated, but a more coherent explanation could not be given \cite{Haase2017}. In a next step, the planar Cu relaxation data were collected \cite{Avramovska2019,Jurkutat2019}, and the expectations were proven wrong again, as these offered a surprisingly simple phenomenology: a universal relaxation rate, independent on material and doping. Previously addressed differences result from the relaxation anisotropy that can differ between systems, but it is always temperature independent. This immediately said that the hitherto adopted shift interpretation must be flawed, i.e., if the relaxation is not enhanced, the shifts must be suppressed \cite{Avramovska2019,Jurkutat2019} where the Korringa relation failed. More recently, after collecting all planar O data another simple phenomenology became apparent, as depicted in Fig.~\ref{fig:fig1}: relaxation and shift are in agreement with a temperature independent but doping dependent pseudogap in a simple metallic density of states that is common to all cuprates \cite{Nachtigal2020,Avramovska2021}, similar to what had been proposed from specific heat data \cite{Loram1998}. 

Very recently, all Cu and O data amassed so far have been compared \cite{Avramovska2022} for an unbiased view of the data, and a number of conclusions could be derived. Foremost, the same pseudogap phenomenon governs the Cu shifts for \cperp. However, the behavior of the shifts for \cpara remained mysterious, but certainly points to a two component model, interestingly, with the same hyperfine coefficients.

Below, we try to give a summary of the current status of the NMR data analysis from which we derive reliable elements that can serve for a better understanding of the cuprates. Since the charge sharing measured with NMR gives important family dependences that we also find in the Cu shifts, we will begin with a brief recapitulation of the planar charges.

\subsection{Charge Sharing in the C\lowercase{u}O$_2$ Plane}
\begin{figure}
	\includegraphics[width=0.45\textwidth]{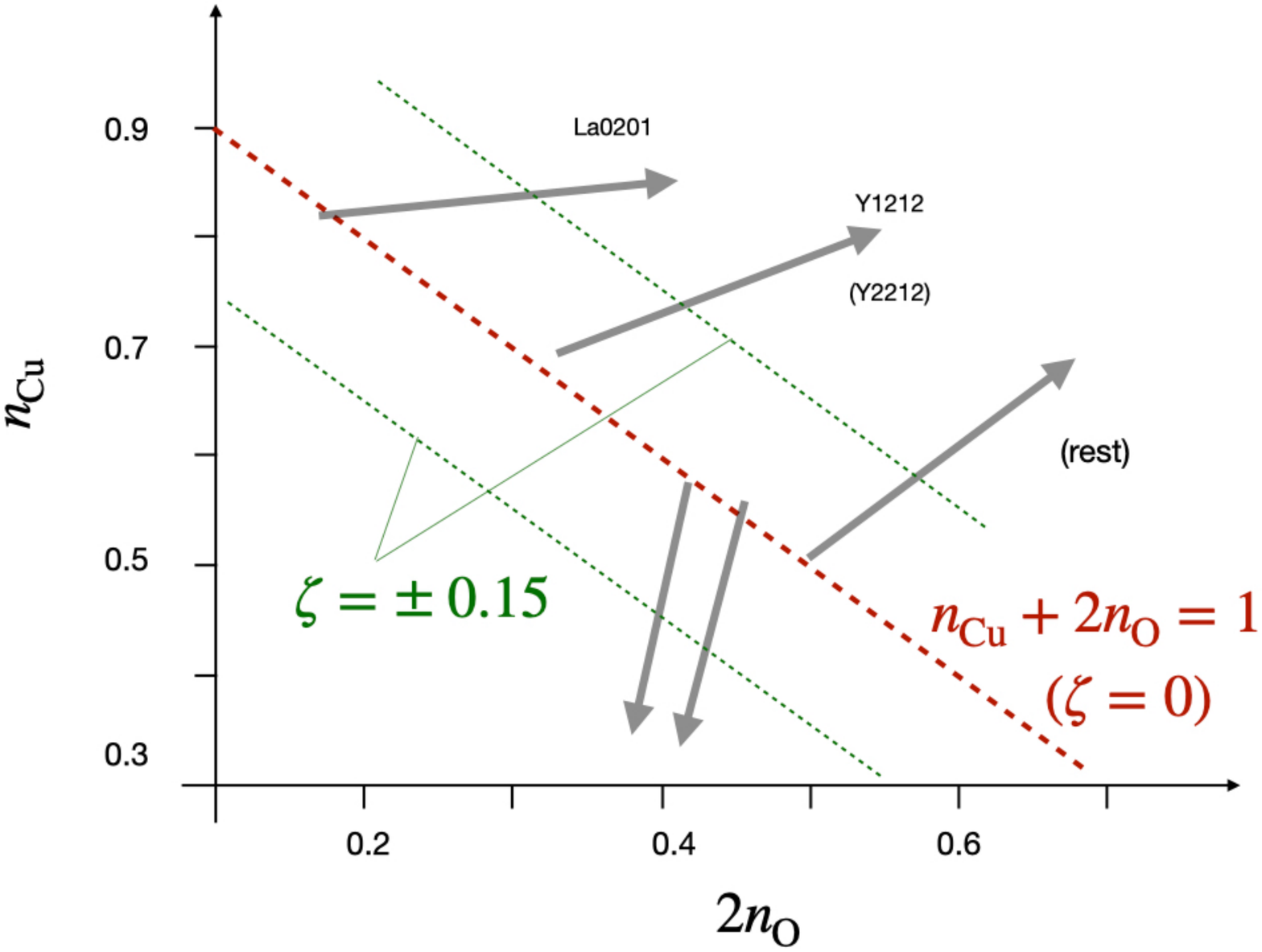}
	\caption{The sharing of the charge in the CuO$_2$ plane between Cu ($n_\mathrm{Cu}$) and O ($n_\mathrm{O}$) as determined with NMR \cite{Jurkutat2014} leads to the NMR doping $\zeta = n_\mathrm{Cu}+2n_\mathrm{O} -1$. Different families can be distinguished, and for the hole doped materials ($\zeta>0$) we find the same families lead to different planar Cu shift anisotropies.}
	\label{fig:fig2}
\end{figure}

It was estimated, early on, that the nominal hole in the Cu $3d(x^2-y^2)$ orbital is responsible for a strong, local electric field gradient \cite{Pennington1989,Martin1995} that splits the planar Cu Zeeman line. Its effect on planar O through hybridization was expected to be much weaker. Later, it was pointed out that the charge sharing between Cu and O appears to be related to \tc \cite{Zheng1995h}, and, after calibrating the NMR quadrupole splitting by means of the electric hyperfine constant, accessible independently from atomic spectroscopy \cite{Haase2004}, it was shown that NMR can measure doping locally. Then, after the first $^{17}$O NMR measurement of electron doped materials were recorded \cite{Jurkutat2014}, it became clear that even the hole distribution in the parent compound could be measured with NMR. The following simple relation holds \cite{Jurkutat2014},
\begin{equation}
1+\zeta = n_\mathrm{Cu} + 2 n_\mathrm{O},
\end{equation}
where $n_\mathrm{Cu}$ and $n_\mathrm{O}$ are the hole contents measured with NMR of planar Cu and O, respectively, cf.~Fig.~\ref{fig:fig2}. The hereby defined parameter $\zeta$ - the doping measured with NMR - agrees well with $x$ in \lsco \cite{Haase2004}, but there are subtle differences between $\zeta$ and what is believed to be the doping (e.g. for \ybco it is observed that at optimal doping $\zeta \approx 0.2$ \cite{Jurkutat2014,Jurkutat2021}).
Most importantly, the sharing of the nominal hole of the parent material ($\zeta = 0$) between Cu and O sets some families apart, and $2n_\mathrm{O}$ was shown to be nearly proportional to the maximum \tc at optimal doping. That is, a phase diagram in terms of $n_\mathrm{Cu}$ and $2n_\mathrm{O}$ carries  greater physical significance, as it can predict the maximum \tc as well as other properties \cite{Jurkutat2019b}. New DMFT calculations \cite{Kowalski2021,Weber2021} confirm this role of the planar O charge.

With regard to the sharing of charge in the parent and as a function of doping, 3 families of hole doped materials are readily distinguished by $n_\mathrm{Cu}$ and $2n_\mathrm{O}$ of the parent, as well as the slope $\Delta_\zeta n_\mathrm{Cu}/\Delta_\zeta (2n_\mathrm{O})$, i.e. how doped holes enter the CuO$_2$ plane, cf.~Fig.~\ref{fig:fig2}. For \lsco the slope is rather small as the doped holes mostly enter the planar O bonding orbital that has the smallest hole content of the cuprates. For the \ybco family of materials the slope is about 0.5 and  for the systems that can have the highest \tc it is slightly larger, cf.~Fig.~\ref{fig:fig2}. We will find below that the planar Cu shift anisotropy is qualitatively different for these 3 groups of materials, as well. 

\subsection{Review of more recent insight}
\paragraph{Planar O NMR}
A metallic density of states common to all cuprates with a temperature independent pseudogap that is set by doping is in agreement with all found literature data (the maximum pseudogap is similar to the exchange coupling $J$) \cite{Nachtigal2020,Avramovska2021}. Excitations across this gap, as the Fermi function opens or closes as a function of temperature, are responsible for the unusual temperature dependences of shift and relaxation, cf.~Fig.~\ref{fig:fig1}. The anisotropies of shift and relaxation are rather independent of temperature and follow from the expected hyperfine coefficient of planar O \cite{Huesser2000}. When the gap is closed one recovers the Korringa relation between planar O shift and relaxation as for a normal metal (the normalized relaxation rate of $1/^{17}T_{1\parallel \sigma}T \approx \SI{0.22}{/Ks}$ for out of plane fluctuations give ${^{17}K}_{\perp\mathrm{c}} \approx \SI{0.2}{\%}$). Note that the original $^{89}$Y NMR data \cite{Alloul1989} agree with this scenario, as well as the specific heat \cite{Loram1998}. The gap varies with doping, but also among families \cite{Nachtigal2020}. At low temperatures, in the presence of a gap, there is small, additional relaxation and shift, perhaps from in-gap states.

\paragraph{Planar Cu relaxation}
The planar Cu relaxation offers a very simple phenomenology \cite{Avramovska2019,Jurkutat2019}. It shows the opening of the superconducting gap at \tc for all superconducting materials, with $1/T_{1\perp}T$ dropping from about \SI{20}{/Ks} at \tc to zero (notable is the absence of a coherence peak). While there is a relaxation anisotropy that varies between about 0.8 and 3.4 for different materials and doping levels, it is temperature independent, so that the rates measured for different directions of the magnetic field are strictly proportional to each other, and we can focus on just one direction (\cperp). 

There is no pseudogap in the planar Cu relaxation, which had been assumed to be dominated by antiferromagnetic spin fluctuations (while it could not be brought in agreement with neutron scattering \cite{Berthier1997}). The temperature dependence of $1/T_{1\perp}$ above \tc  lags behind ideal metallic behavior above about \SI{200}{K}. However, this is true even for the most overdoped systems (even those that do not superconduct), and they share the same $1/T_{1\perp}T$ below \SI{200}{K} with the most underdoped materials for which data are available (doping just below 10\%).  

Note that the corresponding relaxation rates are related to local field fluctuations ($\left<h_{\parallel,\perp}^2\right>$) with the correlation time $\tau_0$ (electronic fast fluctuation limit) \cite{Pennington1989} by,
\begin{equation}\begin{split} \label{eq:relax10}
\frac{1}{T_{1\parallel}}&=\frac{3}{2}\gamma_n^2\cdot 2\left<h_\perp^2\right> \tau_0,\\
\frac{1}{T_{1\perp}}&=\frac{3}{2}\gamma_n^2\left[\left<h_\perp^2\right>+\left<h_\parallel^2\right>\right] \tau_0.
\end{split}\end{equation}
Thus, a rather ubiquitous $1/T_{1\perp}$ and a material dependent $1/T_{1\parallel}$, as found for planar Cu, point to a fairly material and temperature independent $\left<h_\parallel^2\right>$ from electronic spin fluctuations with fixed $\tau_0$, i.e., along the crystal $c$-axis (with $A_\parallel$).

Obviously, we know from the planar O data that such a density of states exists, except that it has a pseudogap for planar O. Then, the pseudogap could have to do with fluctuating field symmetry, and we discuss this below. If we invoke the Korringa relation we find a shift of 0.89\% for $1/T_{1\perp}T=\SI{20}{/Ks}$, about the shift range seen for planar Cu, cf.\@ Fig.~\ref{fig:fig3}. \par\medskip

\paragraph{Planar Cu shifts}
As was noticed in the old picture, the shifts ${^{63}K}_\perp$ are similar to those of planar O for all directions (${^{17}K}_\theta$), a recent analysis gives more details \cite{Avramovska2022}. Thus, ${^{63}K}_\perp(T, \zeta)$ shows the same pseudogap behavior as O. If one inspects all planar Cu data \cite{Avramovska2022} one finds that there are many materials for which ${^{63}K}_\parallel(T, \zeta)$ is also temperature dependent, unlike \lsco. Furthermore, if ${^{63}K}_\parallel(T, \zeta)$ is temperature dependent it shows the same pseudogap behavior as ${^{63}K}_\perp$ and planar O \cite{Avramovska2022}. Note that these large differences, between \lsco and other materials, cannot be caused by changes in the hyperfine coefficients (since ${^{63}K}_\perp(T, \zeta)$ and ${^{17}K}_\theta(T, \zeta)$ are not affected, as well as the universal relaxation). 

This is best seen if one inspects the Cu shift anisotropy of all materials, i.e. in plots of both shifts against each other, ${^{63}K}_\perp(T, \zeta)$ vs. ${^{63}K}_\parallel(T, \zeta)$, as in figure 7 in \cite{Haase2017}, a sketch is provided in Fig.~\ref{fig:fig3}. In such a plot all data points lie on lines with 3 different slopes that reign in certain ranges of temperature or doping. These slopes are defined by,
\begin{equation}\label{eq:slopes}
\kappa = \Delta_{T,\zeta} {^{63}K}_\perp(T, \zeta)/\Delta_{T,\zeta}{^{63}K}_\parallel(T, \zeta),
\end{equation}
and $\kappa_1 \gtrsim 10$, $\kappa_2 = 1$, and $\kappa_3 = 2.5$. By itself, the appearance of 3 different slopes would mean 3 different hyperfine coefficients for different ranges of temperature in a single component picture, an unlikely scenario. The steep slope is not only found for \lsco, as mentioned above, also very different systems show it in certain ranges of temperature. Most systems assume $\kappa_3 = 2.5$ at lower temperatures, and $\kappa_2 = 1$ as a function of doping sets the three families apart.

Since the old hyperfine scenario has essentially removed the planar Cu $3d(x^2-y^2)$ spin from the shift scenario and, in a sense, replaced it by the transferred coupling (from planar O), this complex anisotropy must be caused by the interplay of the metallic spin with that of the $3d(x^2-y^2)$ orbital (see below).

\subsection{Elements of the new picture}
We believe that in an approach to understand the NMR data of the cuprates, one has to focus on the following findings.
 
{\flushleft (1)} There is  the universal metallic density of states in the cuprates outside a doping dependent pseudogap that disappears for high doping levels. The pseudogap affects planar O shifts and relaxation, as well as $^{63}K_\perp$, and it can also affect $^{63}K_\parallel$, but it is not found for planar Cu relaxation. 

{\flushleft (2)} The superconducting gap is only clearly seen in the planar Cu relaxation data that almost behave in a classical way and fall from a universal reduced rate of $1/T_{1\perp}T_\mathrm{c} \sim \SI{20}{/Ks}$ to zero. The relaxation anisotropy is strictly temperature independent, but differs between the materials (0.5 to 3.4).

{\flushleft (3)} The orbital shifts are understood for planar O and for \cperp and Cu. There must be another significant Cu shift for \cpara of $\gtrsim 0.6\%$. We have good reason to believe that it is just spin shift, and not due to orbital currents \cite{Varma2006}.

{\flushleft (4)} The hyperfine coefficient of the $3d(x^2-y^2)$ spins is known, with $A_\parallel < 0$ and $|A_\parallel| \gtrsim 6 |A_\perp|$. Interestingly, it seems to be a reliable fact that $A_\parallel \approx - 4B$ \cite{Huesser2000,Pennington1989}.

{\flushleft (5)} There is the new understanding of the charge sharing and its relation to the maximum \tc. It sets apart \lsco from \ybco, and also from the rest of the materials \cite{Jurkutat2014}. Interestingly, these families differ also in terms of ${^{63}K}_\parallel$.

\subsection{Towards the understanding of the uniform response}
Planar O shift and relaxation with expected anisotropies \cite{Avramovska2021} show that a single metallic spin component, $\chi_2(T,\zeta)$, is behind the temperature ($T$) and doping ($\zeta$) dependences. The density of states is common to all cuprates and the pseudogap is set by doping, so we have a simple description of the spin shift,
\begin{equation}
{^{17}K}_\theta(T,\zeta) = C_\theta \cdot \chi_2(T,\zeta),
\end{equation}
where $C_\theta$ is the orientation dependent hyperfine constant (in the old literature it was $2C$). $\chi_2$ also changes at \tc, but it is not easily discerned in the presence of the pseudogap. Importantly, the Korringa relation holds at all doping levels (in the sense described above), without major influence of electronic correlations.

Then, this electronic spin must couple to the Cu nucleus and contribute to the planar Cu shifts. In the old scenario this is the component that defined the single fluid picture since it dominates the planar O shift and that of Cu for \cperp. So we can approximate the Cu shift by,
\begin{equation}\label{eq:xx1}
{^{63}K}_{\perp}(T,\zeta) \approx 4B\cdot \chi_2(T,\zeta),
\end{equation}
where we assumed $B$ to be isotropic (we use the factor of 4 as in the old scenario), and we will see in the next paragraph that this is indeed the case. Before we do so, we note that since ${^{17}K}_\theta$ and ${^{63}K}_\perp$ approach zero spin shift at low temperatures for all materials, the low temperature total shift must be the orbital shift. This is supported by first principle calculations for both nuclei \cite{Renold2003}. Thus,
\begin{equation}
\chi_2(T\rightarrow 0) \approx 0,
\end{equation}
as for spin singlet pairing.

\begin{figure}
	\includegraphics[width=0.45\textwidth]{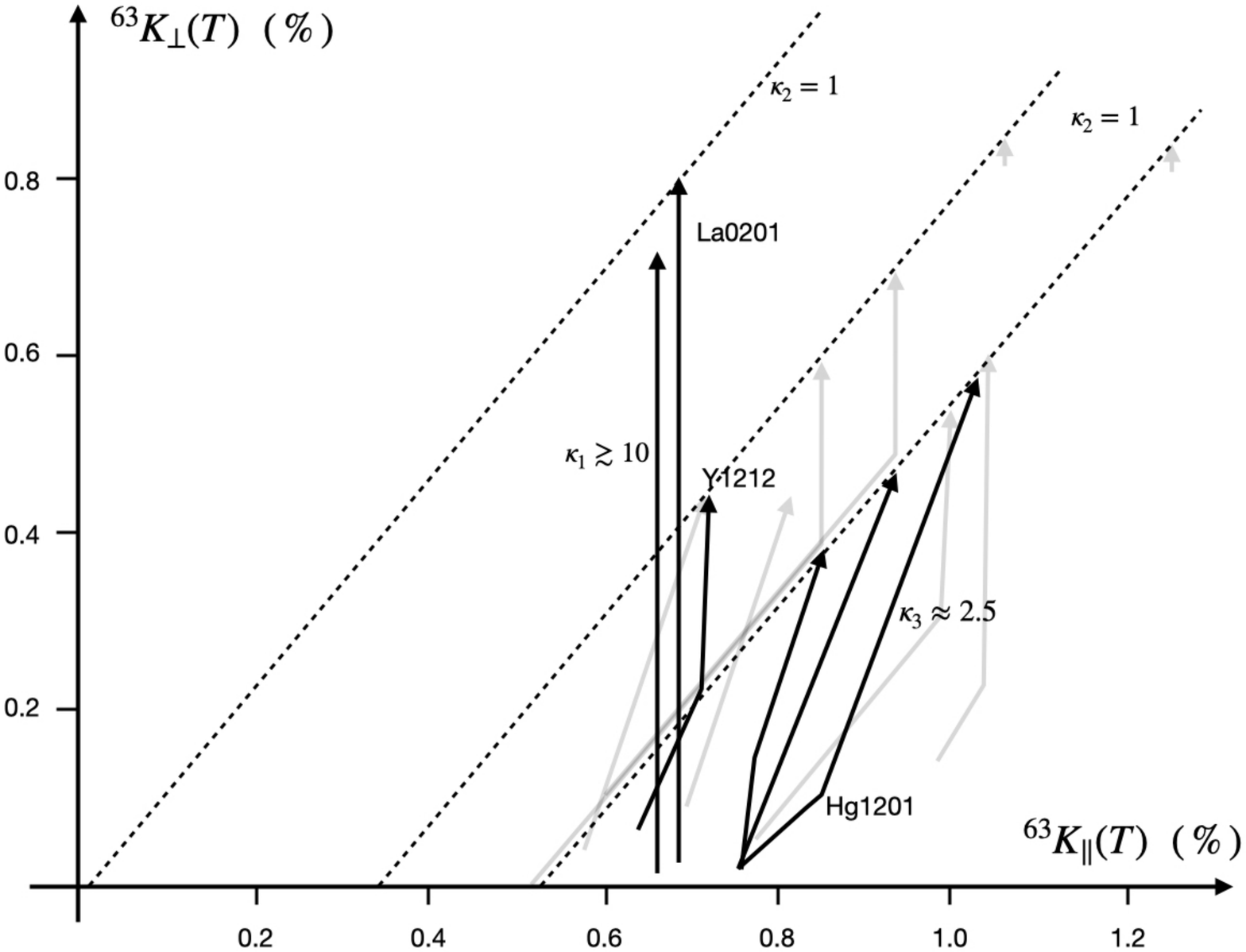}
	\caption{Sketch of the planar Cu shift anisotropy from \cite{Haase2017} for some materials (the full lines will be discussed in the main text); note that the orbital shifts have been subtracted (0.30\% and 0.70\% for \cperp and \cpara, respectively). There are 3 different slopes: the steep one $\kappa_1 \gtrsim 10$, the isotropic shift lines for $\kappa_2 =1$, and the common slope of $\kappa_3 \approx 2.5$. Arrows point in the direction of increasing temperature; the doping increases to the right along $\kappa_2 =1$, or in vertical direction for \lsco. As a function of doping and temperature materials assume one of the 3 slopes, as the two spin components as well as their coupling changes.}
	\label{fig:fig3}
\end{figure}

We now address how this metallic spin affects the Cu shift for \cpara, i.e., the Cu shift anisotropy. We remember that this is not trivial, and there are 3 slopes, defined in \eqref{eq:slopes}, that govern the plot of ${^{63}K}_\perp$ against ${^{63}K}_\parallel$, cf.~Fig.~\ref{fig:fig3}. We note that the lines with a slope $\kappa_2 = 1$ (called the isotropic shift lines in \cite{Haase2017}) are often observed as a function of doping for shifts at higher temperatures, i.e., when the shifts have become largely temperature independent (metal like), they can still vary as a function of doping due to the closing of the pseudogap \cite{Nachtigal2020}. Thus, $\kappa_2 = 1$ is caused by doping dependent changes of $\chi_2$, and a slope of 1 tells us that the associated hyperfine coefficient $B$ in \eqref{eq:xx1} is indeed isotropic, thus $\Delta_\zeta{^{63}K}_\parallel \approx 4B \Delta_\zeta \chi_2$.

Now, we arrive at an important conclusion by inspecting materials with slope $\kappa_2 = 1$. As the temperature is lowered, these (metallic) shifts become eventually temperature dependent, and at that point they assume a different slope ($\kappa_1$ or $\kappa_3$), cf.~Fig.~\ref{fig:fig3}. Thus, there must be another spin component involved that changes as a function of temperature, as well, and we call this component $\chi_1(T)$. This is an important conclusion. Then, with $\chi_2(T,\zeta)$ largely describing ${^{63}K}_\perp$, the hyperfine coupling between the Cu nucleus and $\chi_1$ must be rather anisotropic as this spin acts predominantly for \cpara. Clearly, this can only be the expected $A_{\parallel,\perp}$ from spin in the $3d(x^2-y^2)$ orbital, so we also adopt its sign, and write, 
\begin{align}
{^{63}K}_{\perp}(T,\zeta) &= 4B\cdot \chi_2(T,\zeta),\label{eq:xx10}\\
{^{63}K}_{\parallel}(T,\zeta) &= A_\parallel \cdot \chi_1(T,\zeta)+ 4B\cdot \chi_2(T,\zeta).\label{eq:xx11}
\end{align}
While the equations \eqref{eq:xx10} and \eqref{eq:xx11} may not be exact, they do form a rather good description of the Cu shifts. That is, we recover the old hyperfine scenario, also in agreement with first principle calculations, but with two different susceptibilities (we postulated these before \cite{Haase2017,Avramovska2020}). Note that $\chi_1$ does not depend on doping $\zeta$ for the families with slope 1, but it does so for other materials, as we see below. Therefore, we added the doping dependence for $\chi_1$, as well.
Let us continue the discussion with the cuprates that have slope 1 at high temperatures, in more detail. As the temperature is lowered, we  find the slopes $\kappa_3 = 2.5$ or $\kappa_1 \gtrsim 10$. With other words, $\chi_1(T)$ and $\chi_2(T)$ change together as a function of temperature: $-A_\parallel \Delta_T \chi_1(T) = 3/5 \cdot 4B\Delta_T \chi_2(T)$ and $-A_\parallel \Delta_T \chi_1(T) = 4B \Delta_T \chi_2(T)$ for slope $\kappa_3 = 2.5$ and $\kappa_1 \gtrsim 10$, respectively. One observes in Fig.~\ref{fig:fig3} that at lower temperatures the slope 1 can appear again (lighter lines) with only $\chi_2$ changing, here as a function of temperature. 

Note that we have reason to assume that $A_\parallel \approx -4B$ so that the slopes mean, $\Delta \chi_1 \approx \Delta \chi_2$, $\Delta \chi_1 \approx 0$, and $\Delta \chi_1 \approx 3/5 \Delta \chi_2$ for $\kappa_1 \gtrsim 10$, $\kappa_2 = 1$, and $\kappa_3 = 2.5$, respectively.

Given the discussion above, we conclude that since the steep slope describes \lsco and \ybcoE as a function of temperature and doping, for these systems $\Delta_\mathrm{T,\zeta}\cdot \chi_1(T,\zeta) = \Delta_\mathrm{T,\zeta}\cdot \chi_2(T,\zeta)$. Therefore, the doping dependence of $\chi_1$ seems to be a family characteristic as we discuss further, below.

Now we look at the absolute size of the susceptibilities. ${^{63}K}_\perp$ ranges between 0 and 0.75\% (with an orbital shift of ${^{63}K}_{\mathrm{L}\perp} = 0.30\%$). If we assume the first principle value of ${^{63}K}_{\mathrm{L}\parallel} = 0.70\%$ for the orbital contribution for \cpara, we find that the low-temperature shifts,  ${^{63}K}_\parallel (T\rightarrow 0)$, vary between 0.5\% and 0.8\% for different materials. Even if there were slight variations of the orbital shift, there is significant spin shift $\chi_1$ far below \tc from the $3d(x^2-y^2)$ spin. In order to learn about the size of $\chi_1$ we can look at the isotropic shift lines. These represent lines with $\Delta\chi_1= 0$. We can extend those lines in Fig.~\ref{fig:fig3} to the intersection with $^{63}K_\perp = 0$ to find out about the size of $A_\parallel  \chi_1(\sim \SI{300}{K})$. Then we compare with $A_\parallel \chi_1(T\rightarrow 0)$ which are the low-temperature experimental values for the various families (note that $\chi_2 =0$ on the abscissa). We see that $\chi_1$ increases with temperature by about $ |A_\parallel|\Delta_T \chi_1 \approx 0.2-0.3\%$ for most of the cuprates, but for \lsco it increases by about 0.6\% so that ${^{63}K}_\parallel(T\rightarrow 0) \approx 0$. \par\medskip 

So far we have discussed the data in terms of the two susceptibilities and we saw that these are correlated. This is expected for coupled spins, as a magnetic field acting on the electronic spin $\erww{S_1}$ will also induce a response at the second spin $\erww{S_2}$ if both spins have a (weak) exchange coupling. The total susceptibility is then given by $\chi_{11}+2\chi_{12}+\chi_{22}$, and eqn. \eqref{eq:xx10} and \eqref{eq:xx11} change into,
\begin{align}
{^{63}K}_{\perp} &= 4B\cdot (b+c)\label{eq:xx20}\\
{^{63}K}_{\parallel} &= A_\parallel \cdot (a+c)+ 4B\cdot (b+c),\label{eq:xx21}
\end{align}
where we used a shortened notation, as before \cite{Avramovska2020}: $a=\chi_{11}, c=\chi_{12}, b=\chi_{22}$. With the assumption of $A_\parallel \approx - 4B$, we have,
\begin{align}
{^{63}K}_{\perp}(T,\zeta) &= 4B\cdot (b+c)\label{eq:xx30}\\
{^{63}K}_{\parallel}(T,\zeta) &=  4B\cdot (b-a).\label{eq:xx31}
\end{align}
We believe that these equations are essential in describing the shifts, and they lead to important conclusions, a few of which were pointed out before \cite{Avramovska2020,Avramovska2022}. Note that we subsume the family dependence in $\zeta$, while it is given by $n_\mathrm{Cu}$ and $n_\mathrm{O}$. The steep slope ($\kappa_1 \gtrsim 10$) means that only the coupling $c(T,\zeta)$ is responding to doping or temperature. The slope $\kappa_2 = 1$ appears if only $b(T,\zeta)$ is affected by doping or temperature. Since we do not observe a horizontal slope, i.e. due to $a$ alone, we conclude that $a$ is rather independent of doping and material. Thus, the variables $b$ and $c$ dominate the whole Cu shift anisotropy in Fig.~\ref{fig:fig3}. With \eqref{eq:xx30} and \eqref{eq:xx31} we can also write,
\begin{equation}\label{eq:shift33}
{^{63}K}_\perp = {^{63}K}_\parallel + 4B (a+c).
\end{equation}

With this, we can understand the shifts. We begin by noting, again, that lines with slope $\kappa_2 =1$ (isotropic shift lines), if present, indicate that only $b$ is changing, or, $a+c = const.$ These lines occur at high temperatures. Where they intersect ${^{63}K}_\perp = 0$ we can read off $4B(a_h+c_h)$, or just $(a_h+c_h)$ for short, the high (h) temperature values. Note that at the lowest temperatures $(b_0+c_0)=0$, but $(a_0+c_0)$ can be quite large as this shift does not disappear. So for a given material with $(a_0+c_0) = - s_0$ we have with $(b_0+c_0) =0$ also $(b_0 - a_0) = + s_0$.

Let us inspect \lsco. We have for high temperatures and large doping $(b_h+c_h) \approx 0.8\%$. And since only $c$ changes (steep slope) at low temperatures $\Delta c \approx 0.8\%$  ($\Delta b \approx 0) $. In view of \eqref{eq:shift33}, if both shifts vanish, $(a+c) = 0$, as well. An isotropic shift line through the origin of the shifts should pass the high temperature points of \lsco, and it does. So this fact supports the choice of the orbital shift \cpara.

As a next example we take \hgryb in Fig.~\ref{fig:fig3}. The high temperature shifts can be found on an isotropic shift line, as only $b$ changes with doping. If we follow the line to its intersection with the abscissa, we find that $(a_h+c_h) \approx -0.5\%$ while at the lowest temperature $(a_0+c_0) \approx - 0.75\%$. We conclude that the change in temperature led to $\Delta a + \Delta c \approx 0.25\%$. In addition, the shift change for \cperp gives $\Delta b + \Delta c = 0.4\%$ and we arrive at $\Delta c \approx 0.24\%$ and $\Delta b \approx 0.16\%$ between $T \approx 0$ and about \SI{300}{K}. The component $a$ did not change with temperature. The high temperature points for the other doping levels show that $\Delta b_h \approx \pm 0.08\%$ for higher (+) and lower (-) doping levels, while $c_h$ has not changed. And since all arrive at the same low temperature shift, the overdoped material exhausts $c_h$ before it reaches the lowest shift and continues from that point with $\kappa_2 =1$, while for the underdoped materials the slope changes to the steep slope at lower temperatures to relinquis the remaining $c(T)$.

It appears that a number of strongly doped materials (further to the right on isotropic shift lines) assume the steep slope at \tc, but they change to $\kappa_2 = 1$ to give up the larger $b$. The materials with the highest \tc seem to prefer to fall with $\kappa_3 = 2.5$ so both changes have to be of given size. There is a trade-off between $b$ and $c$ between the different materials (and thus to achieve the highest \tc). Perhaps the different isotropic shift lines are just a consequence of a different coupling strength. Note that at high temperatures there is no negative shift for \cpara for \lsco, but for \hgryb it is rather large, whereas \ybco ranges in between. Also, the shifts ${^{63}K}_\perp$ appear to be lower for the systems with larger coupling (smaller high temperature $c$). A larger ${^{63}K}_\parallel$ for antiferromagnetically coupled spins may indicate stronger coupling to the metallic spins, as well. While this is likely, we do not know for certain whether the orbital shifts are also slightly material dependent for this direction of the field, so that they could account for some differences, as well.

\lsco and \ybcoE are entirely dominated by the steep slope. For all the other systems, a change from the high temperature $\kappa_2 =1$ slope is initiated by the temperature dependence of the shift, either from \tc or the pseudogap. But what causes the changes in slope at lower temperatures is not clear (maybe it is just demanded by the condensate that $c$ and $b$ have to disappear). With the shift measured at different fields, it seem not likely that the latter plays a role in change of the slope. Clues from other probes or theory would be desirable.

\subsection{Nuclear relaxation}
Planar O relaxation \cite{Nachtigal2020,Avramovska2021} is determined by the pseudogap in the metal density of states, changes at \tc are difficult do discern if the gap is sizable (as for all shifts). Only if the pseudogap is nearly closed, \tc decisively changes the temperature dependence. This is in agreement with the action of the metallic spin that dominates planar O for all directions of the field (according to the anisotropy of the hyperfine coefficient). It has been shown that near \tc even relaxation due to the quadrupolar interaction takes place, but not exclusively \cite{Suter2000}. This, however, does not change the general behavior forced by the pseudogap. There is excess relaxation at temperatures near or below the size of the pseuogap. However, since we do not know the exact description of the pseudogap, or the influence of a gap inhomogeneity, it is currently difficult to discuss the behavior in more detail.

Planar Cu relaxation \cite{Avramovska2019,Jurkutat2019} is not at all affected by the pseudogap and shows almost classical behavior, i.e., $1/T_1T$ plotted as a function of $(T/T_\mathrm{c})$ suddenly drops to zero, starting just below \tc. In the framework of fluctuating local fields \cite{Pennington1989}, cf.~\eqref{eq:relax10}, and a rather material and doping independent $1/T_{1\perp}T_\mathrm{c}$ points to spin with a large hyperfine coefficient perpendicular to the CuO$_2$ plane, thus $A_\parallel$. Furthermore, the correlation time $\tau_0$ can only change significantly at \tc. The fact that at temperatures ($\gtrsim \SI{200}{K}$) $1/T_1$ lags behind being proportional to $T$ shows that these fluctuations differ from those observed at planar O, as one might expect since they come from strongly correlated antiferromagnetic spins that interact with the metallic ones. The material dependent anisotropy can be understood as a change in correlation between two spin components in a simple model that offers perhaps an explanation of the pseudogap, as well \cite{Avramovska2020}. It appears that the Cu spin imposes a correlation with the metallic spin that leads to frustration due to the antiferromagnetic coupling, and the since the planar O nucleus only sees the latter, it must overcome the effective gap to scatter the O nucleus. This restriction does not apply to the Cu nucleus.

\subsection{Conclusion}
A review of essential old assumptions and their weaknesses, together with more recent findings, in particular also involving the analyses of literature data, makes it apparent that NMR demands a two component description of the cuprates. There is a metallic reservoir of excitations, from underdoped to strongly overdoped materials, with a density of states common to all cuprates, but it carries a temperature independent pseudogap set by doping. This metallic spin dominates planar O NMR, but also couples to the planar Cu nucleus. However, it is the intricate interaction of this metallic spin with the strongly antiferromagnetically coupled Cu $3d(x^2-y^2)$ spin that leads to the various effects, the pseudgap in shift and relaxation for planar O, and in the shift for planar Cu, but also the universal Cu relaxation. The interplay of the two components is most easily seen in the  planar Cu shift anisotropy that readily shows the rules of interaction as a function of doping and temperature. For example, the anisotropy jumps between three different slopes caused by the two spins and their coupling, with family dependent characteristics, and optimal doping prefers a fixed slope set by the relative sizes of coupling and polarization, and the maximal \tc maybe related to the coupling, as well.  One can only hope that this new phenomenology finds explanations in theory.

 \section*{Acknowledgements}
 	We acknowledge discussions in particular with Marija Avramovska, Jakob Nachtigal, Stefan Tsankov, as well as with Anastasia Aristova, Robin Guehne, Andreas P\"oppl, Daniel Bandur. We thank B. Fine (Leipzig) and A. Erb (Munich) for their extensive communications.

 \section*{Conflict of interest}
 The author declare no conflict of interest.
\bibliography{../BibLibraries/JH-Cuprate.bib}
\printindex


\end{document}